\documentclass[prl,twocolumn,floatfix,showpacs,superscriptaddress]{revtex4}
\usepackage{amsmath}
\usepackage{graphicx}

\def\gtrsim{\mathrel{\raise.4ex\hbox{$>$}\kern-0.8em\lower.7ex\hbox{$\sim$}}}
\def\lesssim{\mathrel{\raise.4ex\hbox{$<$}\kern-0.8em\lower.7ex\hbox{$\sim$}}}

\newcommand{\beq}{\begin{equation}}
\newcommand{\beqn}{\begin{eqnarray}}
\newcommand{\eeq}{\end{equation}}
\newcommand{\eeqn}{\end{eqnarray}}

\begin{document}

\title{Impurity scattering in strongly correlated metals close to the Mott transition}

\author{Pascal Lederer}

\affiliation{Laboratoire de Physique des Solides, CNRS et Universit\'e
Paris-Sud, B\^atiment 510, F-91405 Orsay, France.}
\affiliation{Departamento de Física, UFPE, 50670-901 Recife, Brasil.}

\author{Marcelo J. Rozenberg}
\affiliation{Laboratoire de Physique des Solides, CNRS et Universit\'e
Paris-Sud, B\^atiment 510, F-91405 Orsay, France.}
\affiliation{Departamento de F\'{\i}sica, FCEN, Universidad de Buenos Aires,
Ciudad Universitaria Pab.I, (1428) Buenos Aires, Argentina.}

\begin{abstract}
This work explores a simple aproximation to describe isolated impurity scattering in a strongly correlated metal. The approximation combines conventional one electron scattering theory and the Dynamic Mean Field Theory to describe strong correlations in the host. 
It becomes exact in several limits, including those of very weak and very strong impurity potentials.
Original electronic structure appears at the impurity site when the impurity potential strength is moderate and the 
host is close to the Mott transition. Our results may provide useful guidance for interpretation of scanning tunneling microscopy experiments in strongly correlated systems.
\end{abstract}

\date{\today}
\maketitle








\vspace{20mm}
{\sl Introduction}.
The physics of dilute impurities in broad band metals  such as Cu or Al is one of the success stories of quantum mechanics in the fifties and sixties \cite{friedel}. By dilute  we mean that physical effect due to interactions between impurities can be negelected, and that the properties described vary linearly with the concentration of impurities. For  broad band metals, electron-electron interactions may be treated as perturbations, because the ratio $E_{el-el}/ E_K$ is small compared to unity. $E_{el-el}\approx$1-2eV is the screened Coulomb interaction between electrons, and $E_K\approx$10eV is the electronic kinetic energy, of the order of $\epsilon_F$, the Fermi energy measured from the bottom of the conduction band.
Historically, the treatment of electron-electron interactions in impure metals has concentrated for a long time on the Kondo problem, for which electron-electron interactions in the host where irrelevant \cite{Kondo}. More recently, renormalizations of the Kondo screening due to correlation effects in the host where also 
considered \cite{fulde,vollhardt}. On the other hand, impurity scattering 
with electron-electron interactions on the impurity site and in the host, have been treated within weak coupling methods in a few simple cases, such as the {\bf Pd}Ni 
system \cite{lederermills}.
In recent years, improvements in local probe techniques such as nuclear magnetic resonance and scanning tunneling microscopy shed new light on the problem of impurity effects in strongly correlated metallic materials, in particular with connection to High Tc superconductivity. For instance, in underdoped superconducting cuprates, non magnetic impurities such as $\rm{Zn}$ or $\rm{Li}$ trigger the appearance of magnetic moments on neighbouring sites \cite{bobroff,davis}. This is a natural phenomenon if the two dimensional CuO plane at metallic doping level is reasonably well described by some variant of the RVB theory \cite{RVB}.

The present paper is dedicated to introduce and discuss the results of
a simple approach to the problem of isolated impurity scattering in a strongly correlated metallic system close to the Mott transition. 

The understanding of the Mott metal-insulator transition 
in a half filled band has evolved, since the early papers by Hubbard \cite{hubbard}, with  Gutzwiller's treatment of strong correlations in narrow band metal \cite{gutz}, and the application of that theory to the half filled band by Brinkman and Rice \cite{brink}. The latter paper showed that close to the critical parameter $U_c$ governing the metal-insulator transition, both the Pauli susceptibility $\chi$ and the effective electronic mass $m^*$ diverge with a constant ratio. This behaviour became understandable thanks to the  Dynamic  Mean Field Theory (DMFT)  \cite{DMFT}. In the simple Hubbard picture of the metal-insulator transition,   the upper band, with doubly occcupied sites,  separates at the transition from  the lower singly-occupied band. Following the DMFT picture, a central peak appears between two well separated ``Hubbard'' bands in the metallic phase. The integrated density $Z$ of the central peak goes to zero with the peak width at the transition. The peak is analogous to the Kondo resonance for the single magnetic impurity problem, except that it is a homogeneous property of the system. In addition, this narrow
feature at the Fermi energy sets a scale for a characteristic temperature $T^*$, which goes to zero at the metal-insulator transition. Approaching the transition, the electronic specific heat coefficient and the Pauli susceptibility diverge at zero $T$ as $1/T^*$ \cite{DMFT}. 

A number of important properties of the impurity in the broad band case  are given by the impurity site Green's function $G_{oo}(\omega)= \sum_{k, k'}G(k, k',\omega)$. All electronic properties of the dilute metal are contained in $G(k,k',\omega)$, which is known exactly for the case when electron-electron interactions may be neglected altogether. $G_{oo}( \omega)$ is entirely computed in terms of the scattering potential $V$, which we will assume for simplicity to be a point scattering one, and the site diagonal host Green's function $G^0_{oo}(\omega)$. 
On the other hand, if one considers the case of a narrow band system, the solution of the DMFT 
equations provides the site diagonal Green's 
function $G^{0,U}_{oo}$ of a strongly interacting host, where $U$ denotes the electron-electron intra atomic ``Hubbard'' interaction potential. This Green's function depends solely on the momentum independent self energy 
$\Sigma(U, \omega)$ and on the non-interacting band structure \cite{DMFT}. 
While the DMFT is exact in the limit of large
spatial dimensions (or lattice connectivity), it can be considered as an 
approximation for a finite dimensional lattice, which is the view that
we adopt here to describe the host.
 
This paper explores the following simple idea: in a certain limit, the impurity properties when the host
system is close to the metal-insulator transition may be approximated by the substitution of the DMFT host
site diagonal Green's function in the expression for the impurity site Green's function. This may be 
considered as a (dynamical) mean field description of the impurity electronic properties. 
This simple approach has several correct limits, including the case of $V\rightarrow 0$ (and any $U$) 
and the case $U\rightarrow 0$ (and any $V$), so one may expect that it may be reasonably accurate
for the general case.

{\sl The model}.
We assume that the host is close to the Mott transition and has a number of equivalent degenerate bands, such as can be found in a cubic crystal. We neglect interband electron-electron interactions.
 
 The Hamiltonian for each band is then:
 \beqn
 \label{model1}
 H&=& H_0 + H_{imp}\\
 H_0&=& \sum_{i,j}t_{ij}c^{\dagger}_{i,\sigma}c_{j,\sigma} +h.c. + U \sum_in_{i,\uparrow}n_{i,\downarrow}\\
\label{model3}
 H_{imp}&=& V (n_{o,\uparrow}+n_{o, \downarrow})
 \eeqn

The site diagonal pure host Green's function is given in the DMFT approximation 
as \cite{DMFT}:
\begin{equation}
\label{dmft}
G_{oo}^{0,U}(\omega) =
\sum_k G^{0,U}(k,\omega)=  \sum_k \frac{1}{ \epsilon_k  -\omega -\Sigma(U,\omega)}                            
\end{equation}
 where $\epsilon_k$ is the host electronic dispersion relation in the limit $U=0$
 
From standard isolated impurity scattering theory \cite{mahan},
the impurity Green's function for $U$=0 is:
\beq
\label{Greenzero}
G^{V,0}(k,k')=G^0(k)\delta_{k,k'} + V\frac{G^0(k)G^0(k')}{1-V\sum_kG^0(k)}
\eeq
 
 The site diagonal impurity Green's function for $U$=0 is given by $G^{V,0}_{oo}=\sum_{k,k'}G(k,k') $:
 \begin{equation}
 \label{zeroimp}
 G^{V,0}_{oo}(\omega)
= \frac{G^0_{oo}(\omega)}{1-VG^0_{oo}(\omega)}
 \end{equation}

{\sl Scattering potential model.}
In the scheme that we propose here, host correlations are taken into account by simply replacing 
the uncorrelated Green's function in Eq.\ref{zeroimp} by the DMFT expression of the correlated one.  The local Green's function at an impurity site 
is thus described by
 \begin{equation}
 \label{correlimp}
 G_{oo}^{V,U}(\omega)= \frac{G^{0,U}_{oo}(\omega)}
{1-VG^{0,U}_{oo}(\omega)}.
 \end{equation}
 
It is important to realize that this expression 
has the correct limiting behaviours that we mentioned before. 
In fact, one easily checks
that in Eq.\ref{correlimp} the Green's function $G_{oo}^{V,U}$ becomes the 
homogeneous system $G_{oo}^{0,U}$ for $V$=0, and the impurity Green's 
function in a normal (ie, non-correlated) host $G^{V,0}_{oo}$ when $U$=0.

It is also worth pointing out that the model  (\ref{model1}-\ref{model3}) 
is {\em not} the exact site diagonal impurity Green's function 
in DMFT scheme \cite{vollhardt}. 
In fact, the derivation of the impurity Green's 
function in a correlated host fully within DMFT (ie, in the limit of large dimensions), 
can be obtained in a straightforward manner using the ``cavity
construction'' \cite{DMFT}. However, in that case one realizes 
that the cavity (ie, the Weiss field) 
of the impurity site {\em exactly} coincides with the cavity of 
the clean homogeneous system.
Therefore, there is no renormalization of the impurity
environment due to the presence of the scattering
potential $V$, or, in other words, the Friedel oscillations on neighboring sites
are suppressed in the limit of infinite dimensions. Thus, that
approach is unlikely to provide
a correct physical description of finite dimensional systems.

Within our scheme, the poles of $G_{oo}^{V,U}(\omega)$ contain all the information about the perturbed local density, the local change of the density of states (DOS), 
and the possible occurrence of bound states. In particular, a bound state at energy $\omega_b$ is determined by the simultaneous equations:
 \beqn
 1-V{\rm Re}[G^{0,U}_{00}(\omega_b)]=0\\
 {\rm Im}[G^{0,U}_{00}(\omega_b)]=0
 \eeqn
 When  the first equation is satisfied for $\omega_b$, and 
${\rm Im}[G^{0,U}_{00}(\omega_b)]$ is small compared to its value in the bulk of the pure metal bands, a resonant state is formed, with width 
$\propto \frac{{\rm Im}[G^{0,U}_{00}(\omega_b)]}
{\partial{\rm Re}[G^{0,U}_{0,0}]/\partial{\omega}|_{\omega=\omega_b}}$. 
 
In principle $V$ is a free parameter. However when describing actual impurities, the effective scattering potential results from the perfect screening constraint. In the language of scattering theory, this constraint results in the well known Friedel sum rule \cite{friedel}, which connects the spherical harmonics phase shifts of the scattered wave function to the charge which has to be screened locally in order to ensure electrostatic equilibrium of the host metal. In the present formulation, the effective scattering potential is such that the integrated displaced density below the Fermi level must counterbalance the difference between the impurity nuclear potential and the host one. If the series of 3d transition elements Ti, Cr, Mn, Fe, Co, Ni are dissolved as dilute impurities in, say, V$_2$O$_3$, each impurity will be described by a potential which attracts below or above the Fermi level the number of states required for electric neutrality of the alloy.

Within the present approach, the electronic structure around the impurity site in real space can be simply obtained by Fourier transformation of an expression similar to Eq.\ref{Greenzero},
where the Fourier transform of $\tilde{G}^0(i-j)$, $G^0(k)$, 
is replaced by the interacting host Green's function $G^{0,U}(k,\omega)$ 
(cf Eq.\ref{dmft}). 
%

{\sl Results.}
It is now straightforward, using the known expression for the host correlated site diagonal Green's function provided by the DMFT approach, to find the impurity electronic structure in the whole range of $V, U$ values. 
\begin{figure}
\centering
\includegraphics[width=7cm,angle=0]{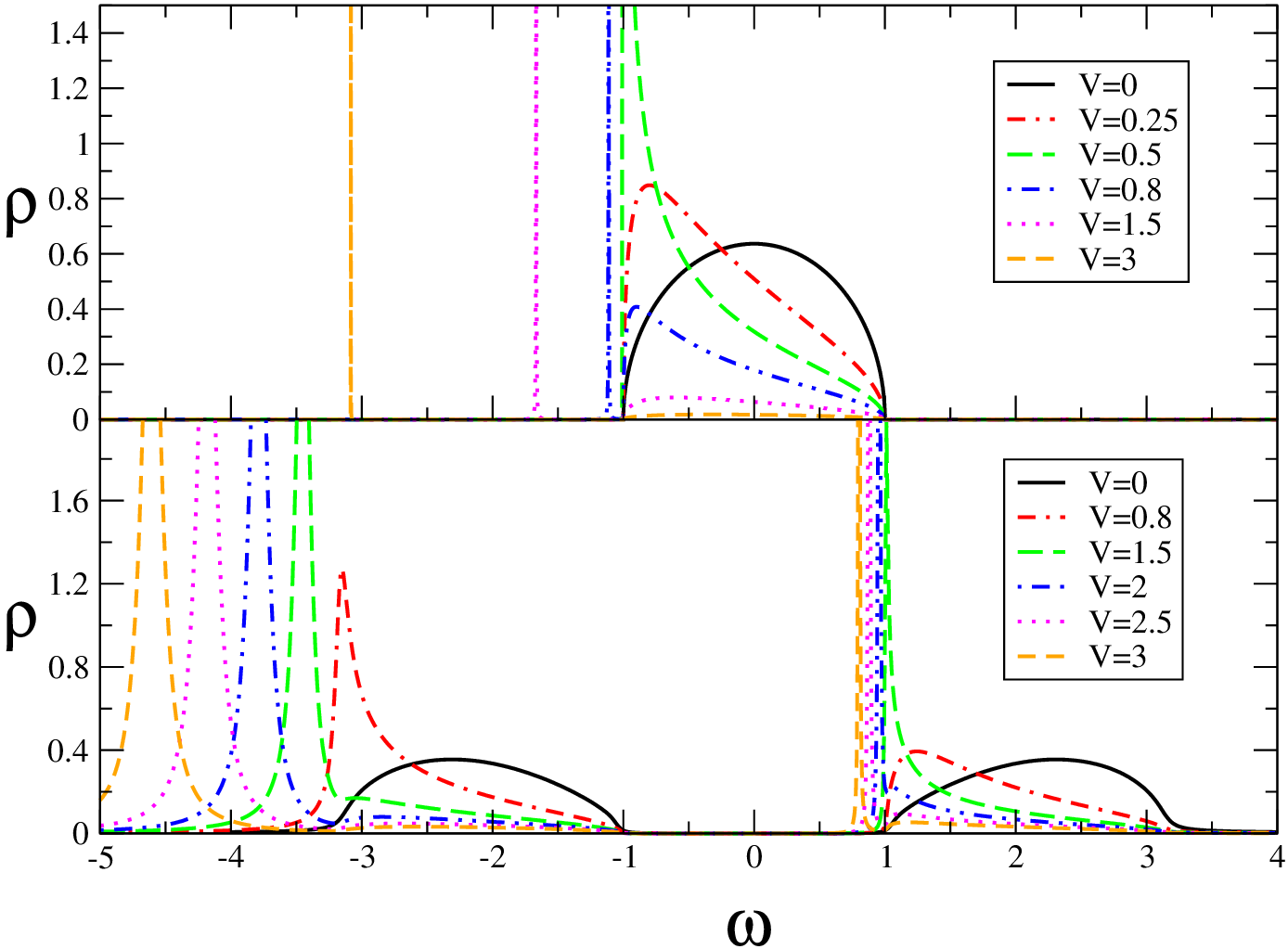}
\caption{
Top panel: Evolution of the impurity density of states $\rho(\omega)$ for
increasing $V$ in a non-correlated host (ie, $U=0$).
Bottom panel: Same quantity when the host is in the Mott insulator state
with $U=4$.
}
\label{densitemodel1}
\end{figure}
We use the Iterated Perturbation Theory to solve for the host
Green's function \cite{DMFT}, adopting a semi-circular non-interacting DOS.
 We begin our discussion of our results with the systematic 
study of the local DOS $\rho(\omega)$ as a 
function of the scattering potential $V$ in two extreme cases. 
In the upper panel of Fig. \ref{densitemodel1} we show 
$\rho(\omega)$ for the case of a metallic non-interacting host (standard Wolf model). In contrast, in the lower panel we show 
the results for the case of an insulating host which is 
well into the Mott state due to strong correlation effects.
The local DOS of the Mott insulator corresponds to
the $V=0$ case (continuous line), where symmetric lower and upper
Hubbard bands split by a gap $\approx U-2D$ can be observed. 
The former case describes the well known results for a non-interacting
metallic host, where the effect of the potential is to initially deform the local DOS 
of the host, shifting spectral weight towards lower energies (since $V<0$), until it becomes 
strong enough to create a bound state out of the conducting band. The condition for the formation of the bound state is $|V| > D$ for a reasonably regular DOS, where $D$ is the half-bandwidth of the metal
host and is taken equal to one, to set the units of the problem.
For a Mott insulating host, as $V$ increases, spectral weight is also shifted to lower energies in both, the upper and the lower Hubbard bands. In addition, there is also a transfer of spectral weight from the upper Hubard band to the lower one. 
Interestingly, as the strength of the scattering potential continues to increase, 
an  impurity resonant state forms simultaneously at the bottom of the upper and the 
lower Hubbard bands. The threshold value for the appearence of the bound states is 
not as clear cut as in the
non-interacting metalic case, but remains of the same order of magnitude. 
It is also worth  noting that the width of the bound state
is very narrow for the upper Hubbard band, while substantially broader 
(and more intense) for the lower Hubbard band.

\begin{figure}
\centering
\includegraphics[width=7.5cm,angle=0]{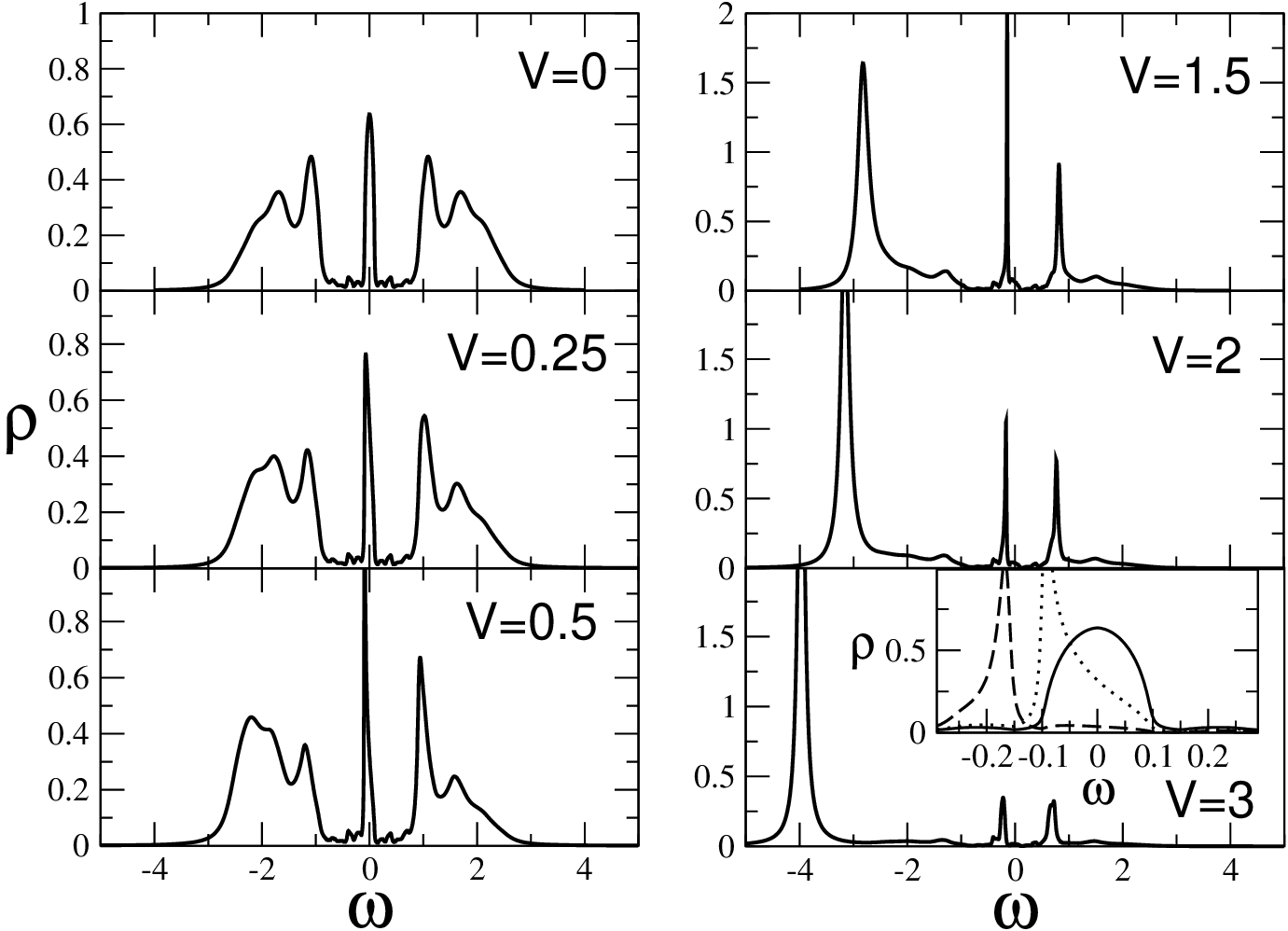}
\caption{The evolution of the 
impurity site $\rho(\omega)$ when the host is close to the Mott transition
with $U=2.9$. 
The panel series show the effect of increasing the strength of the
scattering potential $V$. The bottom panel inset shows 
the detail of the evolution of the mid gap resonance for
$V=0$, 0.5 and 2. 
}
\label{densitemodel3}
\end{figure}

A correlated metal host is well described with DMFT, and the 
clean (ie $V=0$) case is shown in Fig. \ref{densitemodel3} (top 
left panel). The local DOS
shows the characteristic ``Kondo'' peak at $\omega=0$
that corresponds to a renormalized band of heavy mass quasiparticles. The effective bandwidth $ZD$ of this feature is
proportional to its spectral intensity (ie the quasiparticle
residue $Z$) and scales with $U_c-U$, where $U_c \approx 3D$ is
the critical value of the interaction where a metal-insulator
occurs. Spectral weight $1-Z$ is thus shifted from 
low energies to frequencies of order $U$, to form the broad 
lower and upper Hubbard bands. At the transition, the quasiparticle peak becomes infinitely narrow and then disapears, leaving behind
just the
higher energy Hubbard bands (cf Fig.\ref{densitemodel1} lower
panel). It is interesting to observe the systematic effect
of the strength of the scattering potential on the local DOS for this strongly correlated metallic host.
As $V$ is stepped up in absolute value, the DOS is shifted to lower energies in both Hubbard bands, as well  as in the central peak. At the same time a finite amount of spectral weight is transferred from the upper Hubbard band to the lower one, below the Fermi level, allowing a local screening at the impurity site. 
For $|V|$ larger than about $1.5D$, a very sharp resonance appears 
at the impurity site. This sharp resonance carries a spectral
intensity $\sim Z$, ie the value of the quasiparticle residue, 
and thus can be interpreted as a bound state pulled out of the 
narrow renormalized band. This interpretation is further substantiated by the results shown in the inset of the lower right panel, that shows a detail of the evolution of the
DOS near the Fermi energy. We observe that the evolution of
$\rho(\omega)$ is qualitatively similar as that of the non-interacting metal (Fig.\ref{densitemodel1} top panel), but
with the frequency scale renormalized down by a factor of $Z$.
Eventually, when the strength of $V$ is large enough, the quasiparticle band picture near the Fermi energy breaks down and
a sole resonant state at a frequency $\omega \approx V+U/2$ is left.

\begin{figure}
\centering
\includegraphics[width=7.5cm,angle=0]{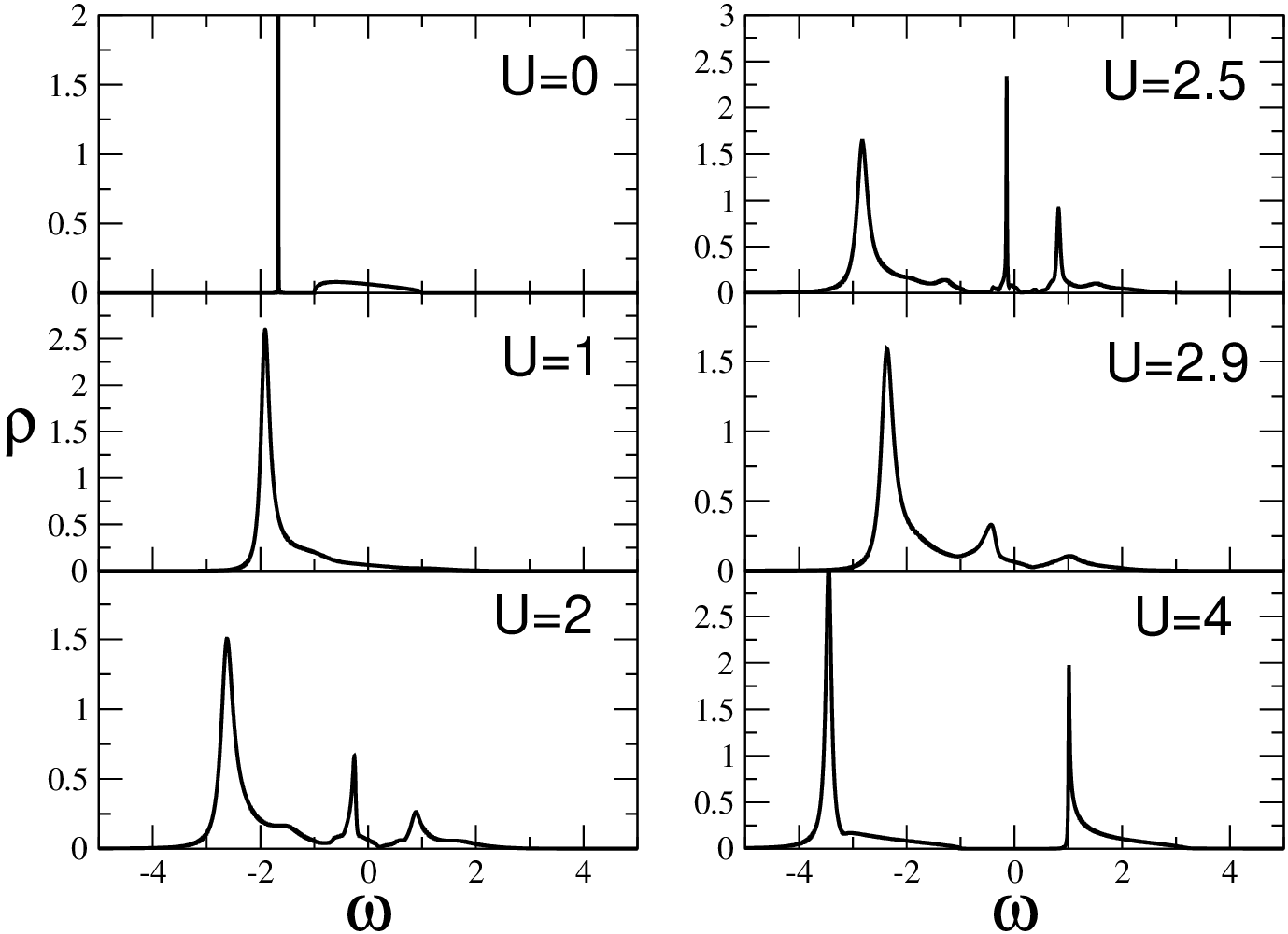}
\caption{Evolution of the impurity density of states
$\rho(\omega)$, for a fixed strong attractive impurity potential $V=1.5$.
The panel series show the effect of increasing the strength of correlations $U$,
as the host is driven across the Mott transition.
}

\label{densitemodel2}
\end{figure}

In Fig.\ref{densitemodel2} we now consider the case of a strongly attractive impurity potential $|V| \geq D$, and study the systematic evolution of the local DOS 
as a function of the electron-electron interaction parameter $U$, as the host
evolves through a metal-insualtor transition. 
For weak $U$, the main effect of el-el interactions is to transform the impurity bound state to a resonant state, due to the tails of the single correlated band. As $U$ is stepped up, the two host Hubbard bands form, and the impurity site develops a multiple-peak structure. There are two resonances associated with the
upper and lower Hubbard bands (with the latter being more prominent as $V$ is assumed attractive), and a central one that is
associated to the central quasiparticle narrow band.
This central feature occurs close to the Fermi energy and
becomes a very sharp resonance at some particular value of $U$, and then broadens again before disapearing as the host becomes
a Mott insulator.
We find that the spectral strength of the mid gap resonance  
at the impurity site scales roughly like the host mid gap 
quasiparticle peak, and both mid gap features vanish together at the Mott 
transition

\begin{figure}
\centering
\includegraphics[width=7cm,angle=0]{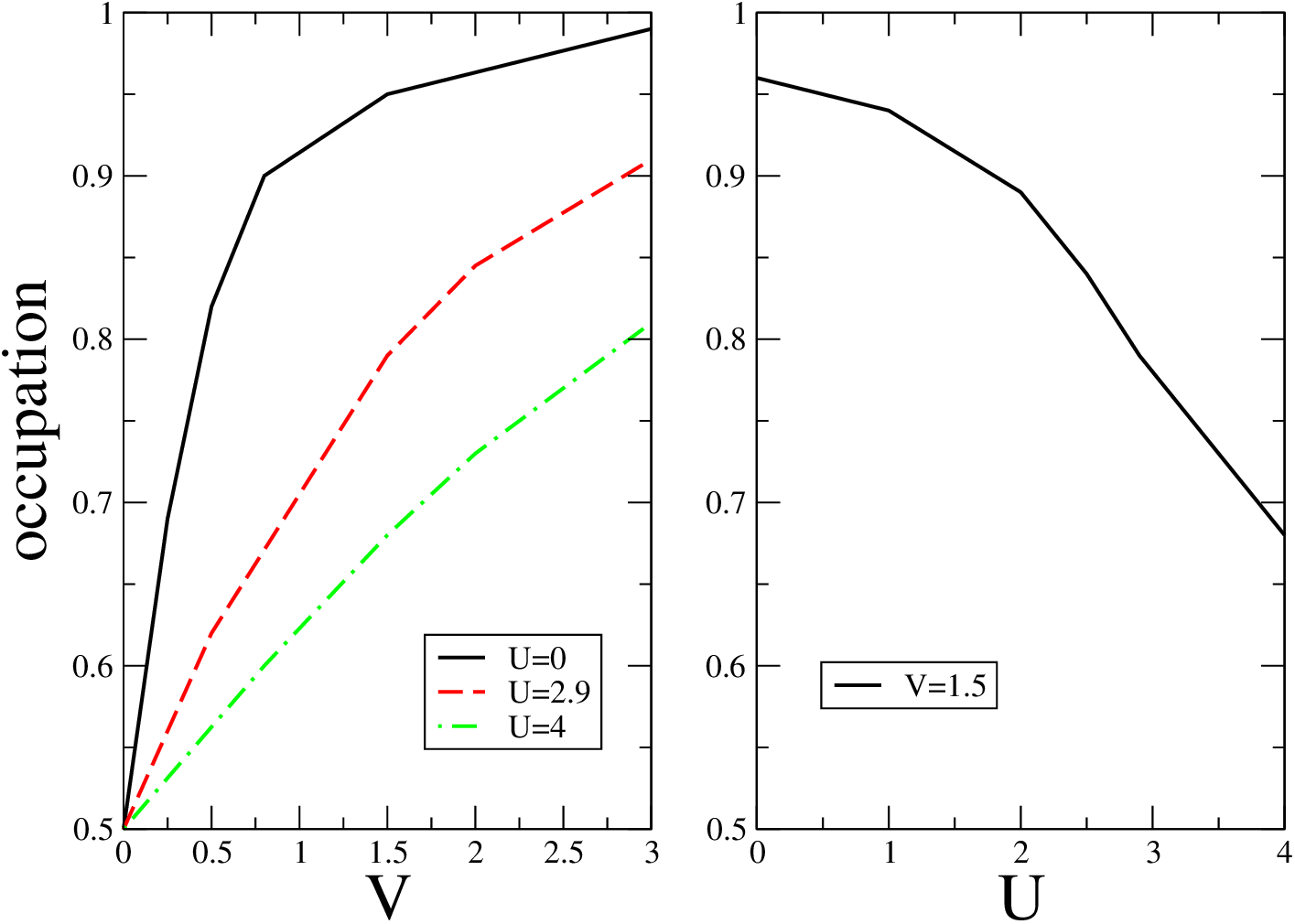}
\caption{The variation of the electronic density at the impurity site
as a function of $|V|$ for various $U$ values (left), and 
as a function of $U$ for various values of $V$ (right). 
}
\label{densitemodel4}
\end{figure}

Finally, Fig.\ref{densitemodel4} exhibits the increase of electronic 
density at the impurity site, as a function of $|V|$ and $U$. As $U$ increases, 
it requires a stronger attractive potential $V$ to screen the same amount of 
impurity nuclear charge. This simply reflects the fact that a large amount 
of states in Hilbert space are pushed up in energy away from the Fermi 
level as $U$ increases, so the amount of states available to screen he 
impurity nuclear potential decreases. 

{\sl Conclusion.}
We believe that the simple approximation described in this work should be confronted with some experimental testing, so as to check its  relevance to actual strongly correlated impure metals, close to the Mott transition. A candidate system to look at would be the V$_2$O$_3$ system, with a small concentration $c$ 
($\approx 10^{-2}$) of Sc, Ti, Cr, Mn, Fe, Co, Ni, or Cu. The measurements of interest are the transport ones, such as  residual resistivity, temperature dependence of the resistivity, thermoelectric power, etc. and the NMR properties, such
as Knight shift, $T_1$. However, perhaps the most clear
cut validation may be obtained through the direct observation of the local electronic density by Scanning Tunneling Spectroscopy or  precision photoemission measurements.

We thank F. Piéchon for useful discussions. MJR acknowledges support from 
ECOS-Sud program.

\end{document}